%% file: ICASSP_2024_main.tex
\documentclass{article}
\usepackage{spconf}

\include{use_packages}
\include{my_def_preample}
\include{def_abr}

\def\BibTeX{{\rm B\kern-.05em{\sc i\kern-.025em b}\kern-.08em
    T\kern-.1667em\lower.7ex\hbox{E}\kern-.125emX}}
\title{IRREGULARITY-AWARE BANDLIMITED APPROXIMATION FOR GRAPH SIGNAL INTERPOLATION}
\name{Darukeesan Pakiyarajah, Eduardo Pavez, Antonio Ortega}
\address{University of Southern California, Los Angeles, CA, USA}
\begin{document}

\ninept
\maketitle
\begin{abstract}
In most work to date, graph signal sampling and reconstruction algorithms are intrinsically tied to graph properties, assuming bandlimitedness and optimal sampling set choices. However, practical scenarios often defy these assumptions, leading to suboptimal performance. In the context of sampling and reconstruction, graph irregularities lead to varying contributions from sampled nodes for interpolation and differing levels of reliability for interpolated nodes. 
The existing \gls{gft}-based methods in the literature make bandlimited signal approximations without considering graph irregularities and the relative significance of nodes, resulting in suboptimal reconstruction performance under various mismatch conditions. In this paper, we leverage the \gls{gft} equipped with a specific inner product to address graph irregularities and account for the relative importance of nodes during the bandlimited signal approximation and interpolation process. Empirical evidence demonstrates that the proposed method outperforms other \gls{gft}-based approaches for bandlimited signal interpolation in challenging scenarios, such as sampling sets selected independently of the underlying graph, low sampling rates, and high noise levels.
\end{abstract}
%
% \vspace{-1ex}
\section{INTRODUCTION}
% \vspace{-1ex}
\label{sec:intro}
\glsresetall

The first step in traditional graph signal sampling and reconstruction is defining a graph and its associated graph operator \cite{tanaka20}. The graph plays a crucial role in representing the underlying data \cite[Ch.~4]{ortega22}. Thus, the processes of sampling and reconstructing graph signals are deeply interconnected, and most of the existing reconstruction methods assume that the signals are band-limited~\cite{anis14,chen15, pesenson08} or smooth~\cite{Tanaka20b_smooth, Bai19} with respect to 
 the \gls{gft} defined using the chosen graph operator. 
 % In \cite{chen15}, the authors proposed a bandlimited model for the sampling and reconstruction of graph signals, which has significantly influenced subsequent research. Notably, many sampling set selection methodologies in the literature, such as those presented in works such as \cite{anis16, Tsitsvero16, jayawant22}, are fundamentally grounded in the principles of bandlimited reconstruction. 
The bandlimited model is employed for sampling set selection in most of the literature \cite{anis16, Tsitsvero16, jayawant22}.
 
 %Additionally, \cite{chen15b} introduced a reconstruction approach based on minimizing the total variation of the signal, leveraging the innate smoothness priors inherent in the signal. 
 
In most cases, even if a graph is selected given its favorable properties, there is no guarantee that the observed signals will be bandlimited.
Moreover, in some cases, the sampling set might be predetermined, or its selection might be based on criteria unrelated to the bandlimited condition. For instance, consider a sensor network employed to measure environmental data. The placement of these sensors across the geographical area is often irregular, influenced by factors like the significance of measurements in various regions, communication costs, and physical access constraints. Moreover, in certain scenarios, only a subset of sensors might be chosen to report their readings, either through random selection or based on specific signal characteristics. 
In such instances, caution is necessary when reconstructing using bandlimited approximations. 
Specifically, these graph structures often feature irregularities with varying node densities across different parts of the network. Some of the sampled nodes hold greater importance for interpolation than others, and the number of samples used influences the reliability of interpolation. Furthermore, traditional reconstruction methods that approximate smooth signals as bandlimited signals for reconstruction fail to consider these relative importance factors.
%are not robust to model mismatches and noise.

 % In \cite{girault18}, a novel concept emerged: an irregularity-aware \gls{gft} for arbitrary Hilbert spaces equipped with an inner product.  This innovation enabled the design of \gls{gft}s that consider the relative significance of graph nodes. The selection of the inner product found utility in various applications: 
 %\cite{girault18} and \cite{girault20} employed it to tackle irregularities in signal sampling for continuous signal energy estimation; 
 % \cite{pavez20} leveraged it to handle irregularities in 3D point cloud geometry, leading to the development of a hierarchical \gls{gft} for point cloud attribute compression; and \cite{Lu20} and \cite{fernandez23} utilized it to consider the relative perceptual importance of individual pixels in image coding. The potential of the inner product extended even further in \cite{pavez22}, where a vertex-partition driven inner product was introduced, enabling the design of a \gls{gft} with spectral folding property for arbitrary graphs, a property previously limited to bipartite graphs before this work.
We focus solely on addressing the interpolation problem within the realm of bandlimited approximation, assuming that the sampling set is already provided. Our work builds upon the foundation of irregularity-aware \gls{gft} \cite{girault18} and its recent extensions to two-channel filterbank design for arbitrary graphs \cite{pavez22}. We leverage the interpretation of the inner product chosen in \cite{pavez22} within the context of sampling and reconstruction. We demonstrate that the corresponding \gls{gft} yields bandlimited approximations that are robust to lack of signal smoothness and presence of noise by taking into account two critical factors: i) the relative importance of nodes and ii) the smoothness of both the sampled and interpolated signals. 
This leads to smaller errors during bandlimited approximation for nodes crucial for interpolation, resulting in more reliable interpolated data. Furthermore, we demonstrate that employing spectral folding \gls{gft} results in a simpler reconstruction formula that does not require inverse terms and eliminates the need for bandwidth estimation, a challenge encountered in other \gls{gft}-based interpolation methods \cite{jayawant23}.

As an application, we consider a sensor network assigned to monitor spatially smooth environmental data. Within this network, the signals demonstrate spatial smoothness and, thus, are indirectly smooth with respect to the graph structure, but this smoothness does not necessarily translate into guaranteeing bandlimited graph signals~\cite{Lee13}. In this context, relying solely on bandlimited approximations without considering irregularities results in larger errors, particularly when the sampling set is chosen independently of the underlying graph structure. In our experiments, we consider random sampling and approximately spatially uniform sampling sets, selected without considering the underlying sensor network graph. The empirical results provide clear evidence that our proposed method is robust to model mismatches when interpolating graph signals under bandlimited approximations, compared to other \gls{gft}-based interpolation methods.

% The rest of the paper is organized as follows: In \cref{sec:review}, we briefly review the fundamentals of the irregularity-aware \gls{gft} and the spectral folding \gls{gft}. In \cref{sec:recon}, we present the proposed interpolation method. In \cref{sec:experiments}, we provide empirical evidence to show the effectiveness of the proposed bandlimited approximation for interpolation and conclude with future directions in \cref{sec:conclusions}.
% \vspace{-1ex}
\section{($\Mm,\Qm$)-GFTs and Spectral Folding}
% \vspace{-1ex}
\label{sec:review}
%First, we review the fundamentals of \gls{gsp}. 
We consider a finite undirected graph $\Gc=(\Vc,\Ec)$ with $N$  vertices, where $\mathcal{V}=\{1,2,\cdots,N\}$ is the vertex set, and $\Ec \subset \Vc \times \Vc$ is the edge set. A graph signal is a function $x: \Vc \to \mathbb{R}$, and the vector representation is given by $\xv = [x_1, x_2, \hdots, x_N] \tr$, where $x_i$ is the signal value at vertex $i \in \Vc$. The adjacency matrix is a non-negative symmetric matrix $\Wm=(w_{ij})$, where $w_{ij}$ is the weight of the edge between vertex $i$ and $j$, and $w_{ij}=0$, whenever $(i,j)\notin \Ec$. $\Dm=\mathrm{diag}([d_1, d_2, \cdots, d_N])$ is the degree matrix of the graph, where $d_i = \sum_{j \in \Vc} w_{ij}$ is the degree of the node $i$. The combinatorial Laplacian is $\Lm=\mathbf{D}-\mathbf{W}$, and the normalized graph Laplacian is $\mathbf{\Lc}=\mathbf{D}^{-1/2}\mathbf{L}\mathbf{D}^{-1/2} = \mathbf{I}-\mathbf{D}^{-1/2}\mathbf{W}\mathbf{D}^{-1/2}$.

In the irregularity-aware \gls{gft} proposed in  \cite{girault18},  graphs are represented by a positive semi-definite variation operator $\Mm \succcurlyeq 0$, which measures signal smoothness, in a Hilbert space where the inner product is defined as $\langle \xv, \yv \rangle_\Qm =  \yv\tr\Qm \xv$ with induced $\Qm$-norm $\|\xv\|_\Qm^2 = \langle \xv, \xv \rangle_\Qm$ where $\Qm \succ 0$. The total variation of a graph signal is defined by $\Delta (\xv) = \xv\tr\Mm \xv$. Then, the $(\Mm, \Qm)$-GFT basis vectors are the columns of $\Um = [\uv_1, \hdots \uv_N]$, which solves the generalized eigendecomposition problem:
\begin{equation}\label{eq:gen_eig}
    \Mm \uv_k = \lambda_k \Qm \uv_k,
\end{equation}
where $0 \leq \lambda_1 \leq \hdots \leq \lambda_N$.
%The set of generalised eigenvalues of a graph is denoted by $\sigma (\Mm, \Qm)$.  
The generalized eigenvectors are $\Qm$-orthogonal (i.e., $\|\uv_i\|_\Qm=1, \: \forall i \in \Vc$, and $\langle \uv_i, \: \uv_j \rangle_\Qm =0$ for all $i\neq j$). With the $(\Mm, \Qm)$-GFT basis, a graph signal $\xv$ is given by
\begin{equation}
    \xv = \sum_{i=1}^N \langle \xv, \: \uv_i \rangle_\Qm \uv_i = \Um \hat{\xv},
\end{equation}
where $\hat{\xv}$ is the $(\Mm, \Qm)$-GFT of $\xv$, with $\hat{x}_i = \langle \xv, \: \uv_i \rangle_\Qm $. The matrix form of this GFT is $\hat{\xv}=\Um\tr\Qm\xv$, while the inverse transform is $\xv=\Um \hat{\xv}$, since $\Um \Um\tr \Qm = \Id$. 

Next, we define the spectral folding property and the spectral folding $(\Mm,\Qm)$-GFT proposed in \cite{pavez22}.
%A linear operator $\Hm$ is a spectral graph filter if there is a function $h: \sigma(\Mm, \Qm) \to \mathbb{R}$ so that $\Hm=\Um h(\Lam)\Um\tr\Qm = h(\Zm)$, where $\Lam = \text{diag}(\lambda_1, \hdots \lambda_N)$ and $\Zm=\Qm^{-1}\Mm=\Um \Lam \Um\tr\Qm$.
\begin{definition} [Spectral folding \cite{pavez22}] \label{def:sf}
Given a graph $\Gc$ with variation operator $\Mm \succcurlyeq 0$, inner product $\Qm \succ 0$ and a partition $\Sc=\{1,\hdots,|\Sc|\}$, $\Sc^c =\Vc \setminus \Sc$. The $(\Mm, \: \Qm)$-GFT has the spectral folding property, if for all $(\uv, \lambda)$ generalized eigenpairs, then $(\Jm\uv, \: (2 - \lambda))$ is also a generalized eigenpair, that is:
\begin{equation}
    \Mm\uv = \lambda \Qm \uv \iff \Mm\Jm\uv = (2 - \lambda)\Qm\Jm\uv,
\end{equation}
where $\Jm$ is a diagonal matrix with diagonal entries given by
\begin{equation}\label{eq:Jmat}
    \Jm_{i,i} = \begin{cases}
        1 & \text{\normalfont{if} } i \in \Sc \\
        -1 & \text{\normalfont{if} } i \in \Sc^c
    \end{cases}.
\end{equation}
\end{definition}
 
\begin{theorem} [Spectral folding $(\Mm,\Qm)$-GFT \cite{pavez22}] \label{the:sfGFT}
    Given a graph $\Gc$ with a variation operator $\Mm \succcurlyeq 0$, and a partition $\Sc = \{1,\hdots |\Sc|\}$, $\Sc^c=\Vc \setminus \Sc$, for which $\Mm_{\Sc\Sc}$ and $\Mm_{\Sc^c\Sc^c}$ are invertible, the corresponding $(\Mm, \Qm)$-GFT has the spectral folding property iff the inner product $\Qm$ is chosen as 
    \begin{equation}\label{eq:def_Q}
        \Qm = \begin{bmatrix}
            \Mm_{\Sc\Sc} & \mathbf{0} \\
            \mathbf{0} & \Mm_{\Sc^c\Sc^c}
        \end{bmatrix}.
    \end{equation}
\end{theorem}
Due to the limited space, we do not present all the properties of spectral folding $(\Mm,\Qm)$-GFT, and the interested reader is referred to~\cite{pavez22}. The relevant properties of this $\Qm$-GFT used in this work are described in \cref{sec:recon}.

\section{PROPOSED INTERPOLATION METHOD USING SPECTRAL FOLDING GFT}
% \vspace{-1ex}
\label{sec:recon}
% \begin{figure}[t]
%     \centering
%     \includegraphics[width = 0.5 \textwidth]{Figures/fig4.eps}
%     \vskip -3ex
%     \caption{Graphs associated with the sensor network: (a) k-nearest neighbour graph constructed for the sensor network for $k=10$, (b) The graph showing the sampled vertices and the connections between them.}
%     \label{fig:sensor_graph}
% \end{figure}

 This paper aims to address the signal interpolation problem under bandlimited approximations. Specifically, given a sampled signal $\xv_\Sc$ defined on a subset of nodes, $\Sc$, of the graph $\Gc$, our goal is to estimate the signal values at the nodes in $\Sc^c$, the complement set of $\Sc$. We consider the combinatorial Laplacian of the graph $\Lm$ as the variation operator and the $(\Lm, \Qm)$-GFT defined in ~\cref{the:sfGFT} with $\Sc$ and $\Sc^c$ chosen as partitions. 
 %as in Thm.~\ref{the:sfGFT}. 
 Note that because of spectral folding \gls{gft}, we have $0\leq \lambda \leq 2$. For our theoretical derivations, we assume that $|\Sc|<|\Sc^c|$, which applies to the context of sampling and reconstruction in general. Next, we define the Paley-Wiener space of $\omega$-bandlimited signals as follows \cite{pesenson08}:
\begin{equation}
    PW_{\omega}(\Gc)=\lbrace \fv : \hat{\fv}(\lambda)=0, \: \forall \lambda > \omega \rbrace \footnote{Here, $\lambda$ are the eigenvalues with respect to $(\Lm,\Qm)$-GFT}.
\end{equation}
 Let $\lambda_r$ be the largest eigenvalue smaller than $1$. We consider the Paley-Weiner space with $\lambda_r$ as the cut-off frequency, $PW_{\lambda_r}(\Gc)$, to harness the spectral folding property for deriving our solution.
 % We make this choice of cutoff frequency to effectively retain the spectral information of the signal with respect to the spectral folding $(\Lm,\Qm)$-GFT and to leverage the spectral folding property for deriving our solution. 
  %with our approach.
 For any $\xv^{bl} \in PW_{\lambda_r}(\Gc)$, we can represent the \gls{gft} coefficients of $\xv^{bl}$ as $\hat{\xv}^{bl}=\begin{bmatrix} \hat\xv_r^{bl\: \top} &\mathbf{0}_{1 \times (N-r)} \end{bmatrix}\tr$, where $\hat\xv_r^{bl} = [\hat x_1,\hdots,\hat x_r]\tr$. This leads to the bandlimited GFT basis representation of $\xv^{bl}$ given by 
\begin{equation}\label{eq:gft_basis_rep}
    \xv^{bl} = \Um_{\Vc \Rc}\hat{\xv}^{bl}=\begin{bmatrix}
        \Um_{\Sc \Rc} \\ \Um_{\Sc^c \Rc}
    \end{bmatrix}\hat{\xv}_r^{bl},
\end{equation}
where $\Rc=\{1,\hdots, r\}$. It is worth noting that with this choice of bandlimited subspace, 
 %we consider all $\lambda$ values less than $1$. This means that 
 bandwidth estimation, a computationally expensive step in other \gls{gft}-based reconstruction methods \cite{jayawant23}, becomes unnecessary.
\vspace{-2ex}
\subsection{Interpretation of the $\Qm$-norm}
\label{sec:innerproduct}
 % where $\Sc$ and $\Sc^c$ denotes the sampled vertices and its complement respectively. 
%First, consider a $(\Lm, \Qm)$-GFT basis vector $\uv = [\uv_\Sc\tr \: \uv_{\Sc^c}\tr]\tr$ corresponding to the eigenvalue $\lambda \:(0 \leq \lambda \leq 2)$ \cite{pavez22}. Using \eqref{eq:gen_eig} and \eqref{eq:def_Q}, we have that
% \begin{align}
%     \Lm_{\Sc\Sc^c}\uv_{\Sc^c} &= (\lambda-1)\Lm_{\Sc\Sc}\uv_{\Sc} \\
%     \Lm_{\Sc^c\Sc}\uv_{\Sc} &= (\lambda-1)\Lm_{\Sc^c\Sc^c}\uv_{\Sc^c}.
% \end{align}
% For $\lambda \neq 1$, this is equivalent to 
% \begin{align}
%     \uv_{\Sc} &= (\lambda-1)^{-1}\Lm_{\Sc\Sc}^{-1}\Lm_{\Sc\Sc^c}\uv_{\Sc^c}  \\
%     \uv_{\Sc^c} &= (\lambda-1)^{-1}\Lm_{\Sc^c\Sc^c}^{-1}\Lm_{\Sc^c\Sc}\uv_{\Sc}.
% \end{align}
% Note that this transform constructs the \gls{gft} basis in a way such that $\uv_{\Sc^c}$ can be computed from $\uv_\Sc$ using the connections going across $\Sc$ and $\Sc^c$, and the connections within $\Sc^c$ and vice versa.
With the $\Qm$-norm chosen in \autoref{the:sfGFT}, for a signal $\xv$ we have 
\begin{equation}\label{eq:normQ}
    \|\xv\|_\Qm^2 = \xv\tr\Qm\xv = \xv_\Sc\tr\Qm_\Sc\xv_\Sc + \xv_{\Sc^c}\tr\Qm_{\Sc^c}\xv_{\Sc^c}. 
\end{equation}
It is worth noting that $\|\xv_\Sc\|_{\Qm_\Sc}^2$ is given by
\begin{equation}\label{eq:normQs}
        \|\xv_\Sc\|_{\Qm_\Sc}^2 = \xv_\Sc\tr\Qm_\Sc\xv_\Sc= \sum_{i\in\Sc}d_i^{\Sc\Sc^c}x_i^2+ \xv_\Sc\tr\tilde{\Lm}_\Sc\xv_\Sc,
    \end{equation}
where $d_i^{\Sc\Sc^c}$ represents the sum of edge weights from vertex $i \in \Sc$ to vertices in $\Sc^c$, and $\tilde{\Lm}_\Sc$ denotes the Laplacian of the graph associated with the vertices in $\Sc$ and their connections. The first term in \eqref{eq:normQs} corresponds to the weighted energy of the signal, with the weights $d_i^{\Sc\Sc^c}$ reflecting the significance of each $x_i$ in the interpolation process. The second term in \eqref{eq:normQs} relates to the smoothness of the signal $\xv_\Sc$ on the sub-graph linked to the vertices in $\Sc$. Similarly, $\|\xv_{\Sc^c}\|_{\Qm_{\Sc^c}}^2$ can be written as
\begin{equation}\label{eq:normQsc}
        \|\xv_{\Sc^c}\|_{\Qm_{\Sc^c}}^2 = \xv_{\Sc^c}\tr\Qm_{\Sc^c}\xv_{\Sc^c}= \sum_{j\in{\Sc^c}}d_j^{\Sc^c\Sc}x_j^2+ \xv_{\Sc^c}\tr\tilde{\Lm}_{\Sc^c}\xv_{\Sc^c},
\end{equation}
where $d_j^{\Sc^c\Sc}$ is the sum of edge weights from vertex $j \in \Sc^c$ to vertices in $\Sc$, and $\tilde{\Lm}_{\Sc^c}$ represents the Laplacian of the graph associated with the vertices in $\Sc^c$ and their connections. The first term in \eqref{eq:normQsc} corresponds to the weighted energy of the signal, where the weights $d_j^{\Sc^c\Sc}$ assign higher importance to nodes from which we can expect more reliable interpolation, i.e., nodes estimated from a larger number of samples. The second term in \eqref{eq:normQsc} relates to the smoothness of the signal $\xv_{\Sc^c}$ on the sub-graph associated with the vertices in $\Sc^c$.

%A similar expression can be derived for $\|\xv_\Sc\|_{\Qm_\Sc}$ using a similar definition of terms.  Thus, from \eqref{eq:normQ} and \eqref{eq:normQs}, we can infer that minimizing $\|\xv\|_\Qm$ is equivalent to enforcing smoothness of $\xv_\Sc$ and $\xv_{\Sc^c}$ on subgraphs associated to vertices in $\Sc$ and $\Sc^c$ respectively.

% Next, 
% Let $\xv^{bl} \in PW_{\omega}(\Gc)$ be the bandlimited approximation of signal $\xv$ and define $\ev = \xv -\xv^{bl}$. Then, 
% \begin{align}\label{eq:approx_error}
%     \|\ev\|_\Qm^2 = \sum_{i\in\Sc}d_i^{\Sc\Sc^c}&(x_i-x_i^{bl})^2 +\sum_{j\in{\Sc^c}}d_i^{\Sc^c\Sc}(x_j-x_j^{bl})^2 \notag \\
%     \qquad &+ \ev_\Sc\tr\tilde{\Lm}_{\Sc}\ev_\Sc + \ev_{\Sc^c}\tr\tilde{\Lm}_{\Sc^c}\ev_{\Sc^c}.
% \end{align}
% It is important to note that minimizing \eqref{eq:approx_error} effectively minimizes the weighted error in the bandlimited approximation, where these weights appropriately consider the significance and reliability of both sampled and interpolated vertices. In addition, minimizing the last two terms in \eqref{eq:approx_error} implies that the error in neighbouring nodes is expected to be similar. Moreover, we emphasize that minimizing $\|\ev\|_2^2$ with respect to the $(\Lm, \Id)$-GFT does not consider this relative importance.

% \subsection{Proposed reconstruction method}
% \label{sec:prop_meth}
% \noindent\textcolor{red}{UNDER EDITING}
\subsection{Proposed solution}
The problem of bandlimited signal approximation and interpolation for a sampled signal $\xv_\Sc$ is formulated as follows:
\vspace{-1ex}
\begin{subequations}\label{eq:opt_prob}
\begin{alignat}{4}
\underset{\yv}{\text{minimize }} &\left\Vert \yv-\begin{bmatrix}
\xv_\Sc \\
\mathbf{0}
\end{bmatrix} \right\Vert_\Qm^2 \label{eq:opt_obj}\\
\text{subject to: }& \yv \in PW_{\lambda_r}(\Gc) \text{ and } \yv_\Sc = \xv_\Sc. \label{eq:opt_cons}
\end{alignat}
\end{subequations}
\begin{theorem}
\label{thm:bl_rec}
The solution to the  optimization problem in \cref{eq:opt_prob} 
has the following closed-form solution:
% \vspace{-1ex}
\begin{align} \label{eq:opt_soln_f}
    \yv = 2\Um_{\Vc \Rc} \Um_{\Sc \Rc}\tr \Qm_\Sc \xv_{\Sc}.
\end{align}
% \vspace{-6ex}
\end{theorem}
% Next, we go into the details of the derivation of the solution for the optimization problem in \eqref{eq:opt_prob}.
%Following this, we go into the details of our primary result. 
The spectral folding property is noteworthy in this result, as it eliminates the need for inverse terms, unlike the solution obtained with $(\Lm, \Id)$-GFT \cite{anis14}.
In \cref{sec:ins_obj}, we justify our choice of the objective function, and in \cref{sec:proof_thm}, we present the proof for \cref{thm:bl_rec}.
\vspace{-2ex}
\subsubsection{Insights into the objective function}
\label{sec:ins_obj}
\vspace{-1ex}
The  objective function defined in \eqref{eq:opt_obj} can be expressed as
\begin{equation}\label{eq:err_vertex}
    \|\ev\|_{\Qm}^2 = \|\yv_\Sc - \xv_\Sc\|_{\Qm_\Sc}^2 + \|\yv_{\Sc^c}\|_{\Qm_{\Sc^c}}^2.
    %\sum_{j\in{\Sc^c}}d_i^{\Sc^c\Sc}y_j^2 + \yv_{\Sc^c}\tr\tilde{\Lm}_{\Sc^c}\yv_{\Sc^c}.
\end{equation}
The constraints in \eqref{eq:opt_cons} imply that $\|\yv_\Sc - \xv_\Sc\|_{\Qm_\Sc}^2=0$, leaving only 
% The first term in the above equation can be further expanded as,
% \begin{align}\label{eq:approx_error}
%     \|\yv_\Sc - \xv_\Sc\|_{\Qm_\Sc}^2 = \sum_{i\in\Sc}d_i^{\Sc\Sc^c}&(y_i-x_i)^2+ \ev_\Sc\tr\tilde{\Lm}_{\Sc}\ev_\Sc,
% \end{align}
% where $\ev_\Sc = \yv_\Sc - \xv_\Sc$. It is important to note that minimizing \eqref{eq:approx_error} effectively minimizes the weighted error in the bandlimited approximation, where these weights appropriately consider the significance of sampled vertices. In addition, minimizing the last term in \eqref{eq:approx_error} implies that the error in neighbouring sampled nodes is expected to be similar in the bandlimited approximation.
the second term in \eqref{eq:err_vertex}, which can be expressed as
\begin{equation}\label{eq:err_vertex2}
     \|\yv_{\Sc^c}\|_{\Qm_{\Sc^c}}^2 =
    \sum_{j\in{\Sc^c}}d_j^{\Sc^c\Sc}y_j^2 + \yv_{\Sc^c}\tr\tilde{\Lm}_{\Sc^c}\yv_{\Sc^c}.
\end{equation} \vskip -1ex
\noindent Note that from \eqref{eq:err_vertex2}, we can interpret the minimization of the objective function as performing regularized smoothing of $\yv_{\Sc^c}$. The second term ensures that  $\yv_{\Sc^c}$ is a smooth graph signal on a subgraph represented by the graph Laplacian $\tilde{\Lm}_{\Sc^c}$  which only considers edges within the nodes in $\Sc^c$. The first term considers weighted energy on interpolated nodes, with the weights $d_i^{\Sc^c\Sc}$ reflecting how well connected is each node in $\Sc^c$ to each node in $\Sc$. This can be viewed as the reliability of the interpolation for each node. 
Here, we emphasize that minimizing $\|\ev\|_2^2$ with respect to the $(\Lm, \Id)$-GFT does not consider this relative importance and smoothness.
\vspace{-1ex}
\subsubsection{Proof of \cref{thm:bl_rec}}
\vspace{-1ex}
% In this work, 
\label{sec:proof_thm}
Our proof relies on the following two results, which will be proved in the next subsections.

% \vspace{-1ex}
\begin{prop} \label{prop:energy_equi}
    For any bandlimited signal $\xv^{bl} \in PW_{\lambda_r}(\Gc)$, $\|\xv^{bl}_\Sc\|_{\Qm_\Sc}^2 = \|\xv^{bl}_{\Sc^c}\|_{\Qm_{\Sc^c}}^2$.
\end{prop}
\vspace{-1ex}
\begin{prop} \label{prop:Q-orth}
    The columns of $\Um_{\Sc\Rc}$ are $\Qm_\Sc$-orthogonal and $\Um_{\Sc\Rc}\tr\Qm_\Sc\Um_{\Sc\Rc}=\frac{1}{2}\Id_{|\Rc|}$.
\end{prop}
\vspace{-1ex}
% The result in \cref{prop:energy_equi} implies that for any $\uv$ with corresponding $\lambda < 1$, it is not possible to have $\uv_\Sc \neq \bs{0}$ and $\uv_{\Sc^c} = \bs{0}$ simultaneously. In other words, none of the columns of $\Um_{\Vc\Rc}$ are localized to the vertices in $\Sc$.

From \cref{prop:energy_equi} and the constraints in \eqref{eq:opt_cons}, we can deduce that the solution to the optimization problem in \eqref{eq:opt_prob} satisfies $ \|\yv_\Sc\|_{\Qm_{\Sc}}^2 = \|\yv_{\Sc^c}\|_{\Qm_{\Sc^c}}^2 = \|\xv_\Sc\|_{\Qm_{\Sc}}^2$. Substituting this back into \eqref{eq:err_vertex} shows that the objective function, when evaluated inside this constraint set, attains the constant value $\|\xv_\Sc\|_{\Qm_{\Sc}}^2$.  Thus, solving the optimization problem in \eqref{eq:opt_prob} boils down to finding $\yv$ that satisfies the constraints in \eqref{eq:opt_cons}. Note that $\yv$ in \eqref{eq:opt_soln_f} takes on the form of the bandlimited signal as defined in \eqref{eq:gft_basis_rep}. Furthermore, $\yv_\Sc$ in \eqref{eq:opt_soln_f} is given by 
\begin{equation}
    \yv_\Sc = 2\Um_{\Sc\Rc}\Um_{\Sc\Rc}^{\top}\Qm_{\Sc}\xv_\Sc = \xv_\Sc,
\end{equation}
where the last equality follows from \cref{prop:Q-orth}, thereby establishing $\yv$ in \eqref{eq:opt_soln_f} as the solution to  \eqref{eq:opt_prob}.
\vspace{-1ex}
\subsubsection{Proof of \cref{prop:energy_equi}}
\vspace{-1ex}
% In this work, 
\label{sec:proof_thm}
% \begin{proof}
First, we define $\xv_{du}^{bl} = \begin{bmatrix}\xv_\Sc^{bl \: \top} & \mathbf{0}\end{bmatrix}\tr$ as the downsampled-upsampled signal, with $\xv_{du}^{bl} = \frac{1}{2}(\xv^{bl}+\Jm\xv^{bl})$, where $\Jm$ is defined as in \eqref{eq:Jmat}. Notably, due to the spectral folding property of ~\cref{def:sf}, the last $r$ columns of $\Um$ are $\Jm\Um_{\Vc\Rc}\Rm_r$ and using $\Qm$-orthogonality of eigenvectors, i.e., $\Um\tr\Qm\Jm\Um_{\Vc\Rc}=\Id_{|\Rc|}$, we show that $\Um\tr\Qm\Jm\Um_{\Vc\Rc}\hat\xv^{bl}_r = \Rm_N \hat\xv^{bl}$, where $\Rm_r$ is the antidiagonal unitary matrix: 
\begin{equation}
    \Rm_r = \begin{bmatrix}
        0 &\hdots & 1\\
        \vdots &\ddots &\vdots \\
        1 & \hdots & 0
        
    \end{bmatrix}_{r \times r}.
\end{equation}
Then the $(\Lm, \Qm)$-\gls{gft} coefficients of $\xv_{du}^{bl}$ are given by
\vspace{-1ex}
\begin{align}
    \hat\xv_{du}^{bl} = 
    % \frac{1}{2}\Um\tr\Qm(\xv^{bl}+\Jm\xv^{bl}) \\
                % &= \frac{1}{2}\Um\tr\Qm(\Um\hat\xv^{bl}+\Jm\Um\hat\xv^{bl}) \\
                \frac{1}{2}(\hat\xv^{bl}+\Rm_N\hat\xv^{bl}). \label{eq:du_gft}
\end{align} \vskip -1ex
\noindent Using the generalized Parseval  identity for arbitrary graph signal Hilbert spaces~\cite{girault18}, we derive the following relations:
\begin{align}
    \|\xv^{bl}\|_\Qm^2 & = \hat\xv^{bl \: \top}\hat\xv^{bl} = \|\hat\xv^{bl}\|_2^2,  \label{eq:ener_rel1}\\
    \|\xv_{du}^{bl}\|_\Qm^2 &=\|\xv_\Sc^{bl}\|_{\Qm_\Sc}^2 = \hat\xv_{du}^{bl \: \top}\hat\xv_{du}^{bl} = \frac{1}{2}\|\hat\xv^{bl}\|_2^2, \label{eq:ener_rel2} \\
    \|\xv_{\Sc^c}^{bl}\|_{\Qm_{\Sc^c}}^2 &= \frac{1}{2}\|\hat\xv^{bl}\|_2^2 = \|\xv_\Sc^{bl}\|_{\Qm_\Sc}^2. \label{eq:ener_rel3}
\end{align}
The last equality in \eqref{eq:ener_rel2} is a consequence of \eqref{eq:du_gft} and $\Rm_N\tr\Rm_N=\Id$.
% \end{proof}
% \begin{proof}
\vspace{-2ex}
\subsubsection{Proof of \cref{prop:Q-orth}}
\vspace{-1ex}
 Leveraging the spectral folding property, it can be shown that the columns of $\Jm\Um_{\Vc\Rc}$ are encompassed within the columns of $\Um$. Therefore, the columns of the matrix $\Vm=\begin{bmatrix}\Um_{\Vc\Rc} & \Jm \Um_{\Vc\Rc}\end{bmatrix}$ are $\Qm$-orthonormal, i.e., $\Vm\tr\Qm\Vm=\Id$. Note that $\lambda_r < 1$ is a necessary condition for this orthogonality to hold, and the proof is omitted here due to space limitations. Then, by expanding the equation $\Vm\tr\Qm\Vm=\Id$ block by block, we obtain the following equations:
\begin{align}
    \begin{bmatrix}
        \Um_{\Sc\Rc} & \Um_{\Sc\Rc}\\
        \Um_{\Sc^c\Rc} & -\Um_{\Sc^c\Rc}
    \end{bmatrix}\tr \begin{bmatrix}
        \Qm_{\Sc} & \mathbf{0}\\
        \mathbf{0} & \Qm_{\Sc^c}
    \end{bmatrix} \begin{bmatrix}
        \Um_{\Sc\Rc} & \Um_{\Sc\Rc}\\
        \Um_{\Sc^c\Rc} & -\Um_{\Sc^c\Rc}
    \end{bmatrix} = \Id, \notag
\end{align}
\vspace{-2ex}
\begin{align}
    \Um_{\Sc\Rc}\tr\Qm_\Sc\Um_{\Sc\Rc}+ \Um_{\Sc^c\Rc}\tr \Qm_{\Sc^c} \Um_{\Sc^c\Rc} = \Id_{|\Rc|}, \\
    \Um_{\Sc\Rc}\tr\Qm_\Sc\Um_{\Sc\Rc}- \Um_{\Sc^c\Rc}\tr \Qm_{\Sc^c} \Um_{\Sc^c\Rc} = \mathbf{0},
\end{align}
which leads to $\Um_{\Sc\Rc}\tr\Qm_\Sc\Um_{\Sc\Rc}=\frac{1}{2}\Id_{|\Rc|}$.     
% \end{proof}
%\vspace{}
 % From \cref{prop:energy_equi}, it can be inferred that $\yv= \begin{bmatrix} \xv_\Sc\tr & \bs{0}\end{bmatrix}\tr$ is not bandlimited, and thus it is not a solution to the optimization problem in \eqref{eq:opt_prob}, which therefore possesses a non-trivial solution for $\yv_{\Sc^c}$ that minimizes \eqref{eq:err_vertex2}. 

 % Furthermore, since $\yv \in PW_{\lambda_r}(\Gc)$, it can be expressed as $\yv=\Um_{\Vc\Rc}\hat\yv_r$, and $\hat\yv_r$ can be obtained from solving the sample consistency constraint in \eqref{eq:opt_cons} which can be expressed as $\Um_{\Sc\Rc}\hat\yv_r = \xv_\Sc$.
% \begin{align}
% \Um_{\Sc\Rc}\hat\yv_r = \xv_\Sc \label{eq:sol_yhat}
% \end{align}
% Further, using \cref{prop:Q-orth}, we can show that $\hat\yv_r=2\Um_{\Sc\Rc}\tr\Qm_\Sc\xv_\Sc$, which leads to the solution in \eqref{eq:opt_soln_f}. 
\vspace{-2ex}
\section{EMPIRICAL RESULTS}
\label{sec:experiments}
\vspace{-1ex}
    \begin{figure*}
        \centering
        \includegraphics[width = 0.98 \textwidth, trim={1.35cm 0.7cm 2cm 0.35cm}, clip ]{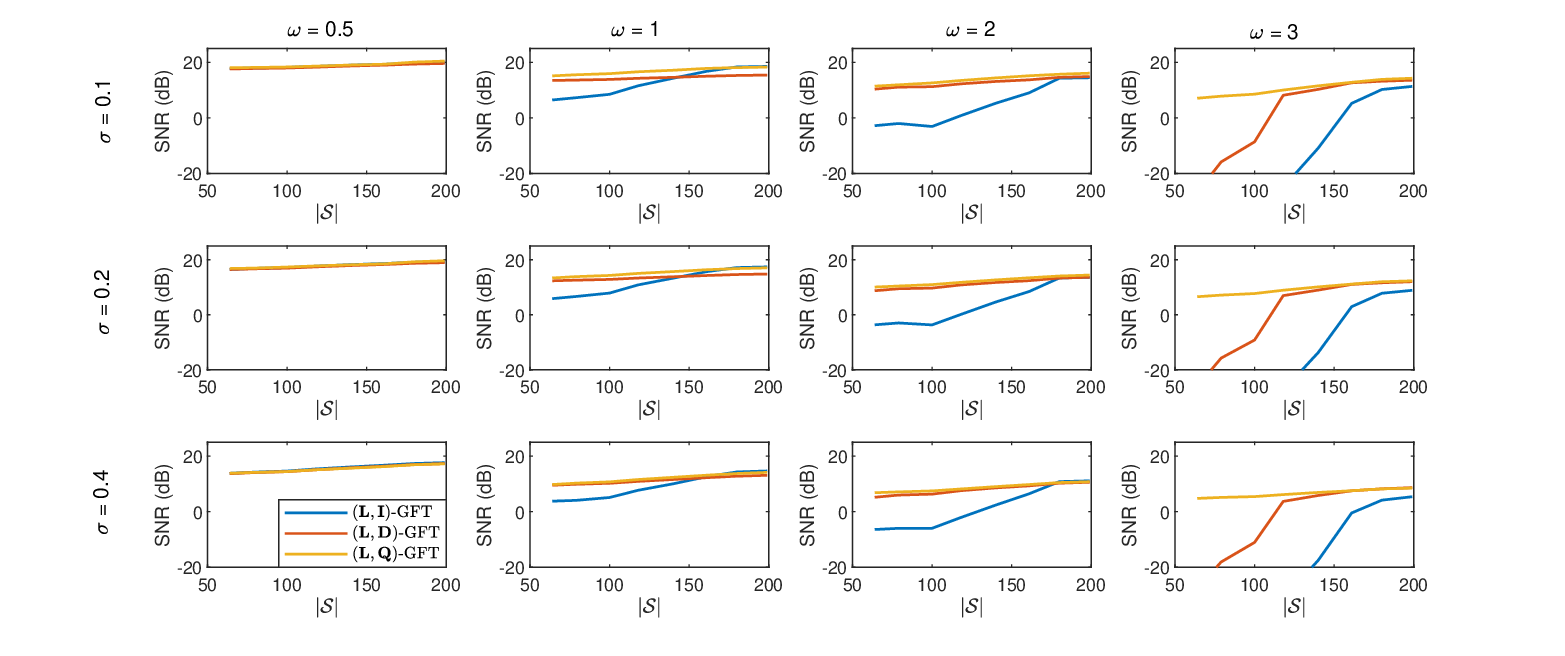}
        \vskip -2ex
        \caption{Comparison of reconstruction error (SNR) for signal $s_n(x_i,y_i)$ for different frequencies $\omega$ and different levels of noise variance $\sigma^2$, for approximately uniform sampling set with varying size $|\Sc|$.}
        \label{fig:exp_sens_netw}
        \vskip -3.5ex
    \end{figure*}
We provide empirical evidence demonstrating the effectiveness of the proposed method compared to existing \gls{gft}-based approaches in bandlimited signal approximation and interpolation within the context of sensor networks. To showcase the potential of the proposed interpolation method, we explore challenging scenarios, including (i) random sampling, (ii) non-bandlimited  signals, and (iii) noisy samples. Our simulations use the Matlab toolboxes GSPBOX \cite{perraudin2014gspbox}, and GraSP \cite{girault17}.

To simulate the process of sampling and interpolation within a sensor network, we define our area of interest as the region $\begin{bmatrix} 0 & 1\end{bmatrix} \times \begin{bmatrix} 0 & 1\end{bmatrix}$ on the Cartesian plane. Within this region, we randomly deploy a total of 500 sensors. The graph $\Gc$ is constructed using a $K$-nearest neighbor algorithm with the parameter $K=8$. Edge weights are determined based on the Gaussian kernel function: $w_{ij} = \exp\left(-d_{ij}^2/2\sigma_d^2\right)$, where $d_{ij}$ represents the Euclidean distance between nodes $i$ and $j$, and $\sigma_d=0.3$. 
For our simulation, we consider a sinusoidal signal that exhibits oscillations solely in the $x$-direction, defined as $s(x,y) = \cos(2\pi \omega x)$. Note that the sinusoidal signal will have $\omega$ oscillations in the $ x$ direction in the region of interest. This is our prototype signal for a spatially smooth signal independent of the underlying graph. The observed signal value at sensor $i$ is denoted as $s(x_i,y_i)$, where $(x_i,y_i)$ corresponds to the sensor's location on the Cartesian plane. 
%We demonstrate the potential of the proposed interpolation methods with challenging scenarios, including (i) random sampling, (ii) higher frequency signals, and (iii) noisy samples.

In the initial experiment, we examine two distinct sampling approaches: i) a randomly selected sampling set and ii) an approximately spatially uniform sampling set, both chosen independently of the graph structure. 
To achieve an approximately spatially uniform sampling set from the available sensors, we partition the Cartesian plane into a grid of equal-area squares and select the sensor closest to the center of each square. The random sampling set and approximately uniform sampling set are denoted as $\Sc_{\text{random}}$ and $\Sc_{\text{uniform}}$ respectively, each containing $100$ nodes. 
We then observe the continuous signal $s(x, y)$ with $\omega=1$ at these sampled vertices and perform interpolation to estimate values at other nodes using  \eqref{eq:opt_soln_f}. Subsequently, we compare the interpolated values to the ground truth signal values and compute the reconstruction error. 
Our method is compared against bandlimited reconstruction using both $(\Lm, \Id)$-GFT and $(\Lm, \Dm)$-GFT. The cutoff frequency for bandlimited reconstruction using $(\Lm, \Id)$-GFT and $(\Lm, \Dm)$-GFT is the $k^{\text{th}}$ eigenvalue of corresponding \gls{gft}, where $k=|\Sc|$. Furthermore, the reconstruction error is computed as the average over $100$ experiments for random sampling. The comparison results, summarized in ~\cref{tab:reconSNR}, demonstrate that the proposed method outperforms other methods regardless of the sampling set selection process. 
Note that using $(\Lm, \Id)$-GFT and $(\Lm, \Dm)$-GFT resulted in a high average reconstruction error for random sampling set selection. 

\begin{table}[]
    \centering
    \begin{tabular}{c c c c}
        \hline \hline
        Sampling set  & $(\Lm, \Id)$-GFT & $(\Lm, \Dm)$-GFT & $(\Lm, \Qm)$-GFT\\ 
        \hline \hline
         $\Sc_{\text{random}}$ & $-12.69$   & $3.35$ & $\bs{14.24}$\\

         \hline 
         $\Sc_{\text{uniform}}$ & $7.37$    &  $19.2$ & $\bs{20.32}$  \\
         \hline 
    \end{tabular}
    \vskip -2ex
    \caption{Comparision of reconstruction error (SNR in dB)  for spatially smooth signal, $s(x_i, y_i)$, with $\omega=1$, with approximately uniform and random sampling sets of size $|\Sc|=100$.}
    \label{tab:reconSNR}
\vskip -4.5ex
\end{table}

In the second experiment, we introduced noise into the signal observations, and we evaluated the reconstruction performance for different smoothness levels of the signal. The noisy signal observations, denoted as $s_n(i)$, were generated as $s_n(i) = s(x_i, y_i) + \eta_i$, where $\eta_i$ follows a normal distribution with mean $0$ and variance $\sigma^2$. We conducted this experiment using approximately spatially uniformly sampled sensors with different sampling set sizes. 
Interpolation was performed for various signal smoothness levels, $\omega = \{0.5, 1, 2, 3\}$, and different noise levels, $\sigma = \{0.1, 0.2, 0.4\}$. As in the previous experiment, our method was compared against $(\Lm, \Id)$-GFT and $(\Lm, \Dm)$-GFT. 
Each experiment was repeated 100 times with different noise realizations, and the average interpolation errors are depicted in \cref{fig:exp_sens_netw}. We observed that under conditions where the signal exhibits smoothness (small $\omega$), all of these methods yielded similar performance. In more challenging scenarios characterized by low sampling rates, non-smooth signals, and high noise levels (see last column of \cref{fig:exp_sens_netw}), the proposed method consistently outperforms the other two GFT-based approaches. 
It is also worth noting that in these scenarios, $(\Lm, \Dm)$-GFT outperforms $(\Lm, \Id)$-GFT.  The improved performance of both $(\Lm, \Qm)$-GFT and $(\Lm, \Dm)$-GFT compared to $(\Lm, \Id)$-GFT underscores the importance of considering graph irregularities when designing GFTs. The consistent performance of $(\Lm, \Qm)$-GFT underscores the value of considering the relative significance of nodes in the context of sampling and reconstruction, which ultimately results in improved bandlimited approximations, especially in challenging scenarios.
\vspace{-2ex}
\section{CONCLUSION}
\label{sec:conclusions}
\vspace{-2ex}
In this work, we considered the problem of bandlimited signal approximation for graph-based signal interpolation. In a graph, certain sampled nodes hold greater importance for interpolation, while some of the interpolated nodes possess more reliable information. Preserving this information during approximation is important. We employed an irregularity-aware \gls{gft} equipped with an inner product $\Qm$ derived from vertex partition to address this challenge. This unique inner product enabled us to consider the relative significance of nodes in bandlimited signal approximation. By harnessing the spectral folding property of the corresponding $(\Lm,\Qm)$-\gls{gft}, we derived a closed-form expression that is less complex compared to the least squares solution obtained using other \gls{gft}s and bandwidth estimation is no longer required. Our empirical results clearly illustrated the benefits of accounting for node importance in bandlimited approximation, particularly when dealing with low sampling rates, graph-independent sampling, and higher noise levels. Further theoretical analysis of employing the spectral folding $(\Lm,\Qm)$-GFT for sampling and reconstruction is left for future work.

% Below is an example of how to insert images. Delete the ``\vspace'' line,
% uncomment the preceding line ``\centerline...'' and replace ``imageX.ps''
% with a suitable PostScript file name.
% -------------------------------------------------------------------------
% \begin{figure}[htb]

% \begin{minipage}[b]{1.0\linewidth}
%   \centering
%   \centerline{\includegraphics[width=8.5cm]{image1}}
% %  \vspace{2.0cm}
%   \centerline{(a) Result 1}\medskip
% \end{minipage}
% %
% \begin{minipage}[b]{.48\linewidth}
%   \centering
%   \centerline{\includegraphics[width=4.0cm]{image3}}
% %  \vspace{1.5cm}
%   \centerline{(b) Results 3}\medskip
% \end{minipage}
% \hfill
% \begin{minipage}[b]{0.48\linewidth}
%   \centering
%   \centerline{\includegraphics[width=4.0cm]{image4}}
% %  \vspace{1.5cm}
%   \centerline{(c) Result 4}\medskip
% \end{minipage}
% %
% \caption{Example of placing a figure with experimental results.}
% \label{fig:res}
% %
% \end{figure}

% \vfill
% \pagebreak

\bibliographystyle{IEEEbib}
% \bibliography{strings,refs}
\bibliography{BibGSP}
\vfill
\end{document}

%% file: use_packages.tex
\usepackage{cite}
\usepackage{url}
\usepackage{textcomp}

\usepackage{array}
\usepackage{multirow}
\usepackage{arydshln}%to have dashlines in tables

\usepackage{amsmath,amssymb,amsfonts}
\usepackage{stfloats}
\usepackage{nicefrac}

\usepackage{graphicx}
\usepackage{subfigure}
\usepackage{xcolor}
\usepackage{tikz,pgfplots}
\usetikzlibrary{arrows}

\usepackage[acronym]{glossaries}

\usepackage{algorithm}
\usepackage{algpseudocode}

\usepackage{caption}
\usepackage{amsmath}
\usepackage{mathtools}

\usepackage{enumitem}
\usepackage{array}
\usepackage{amsthm}
\usepackage[titletoc,toc,title]{appendix}
\usepackage[bookmarksopen=true]{hyperref}
\usepackage[capitalise,nameinlink]{cleveref}

\usepackage{bookmark}

\usepackage{csquotes}
\usepackage{url}

%% file: my_def_preample.tex
\graphicspath{Figures/}
\newcommand{\tr}{^\top}			% transpose of a matrix
\newcommand{\bs}[1]{\boldsymbol{#1}}	% bold symbol
\allowdisplaybreaks[4]					% allow breaks for equations
\long\def\comment#1{}

\newfont{\bbb}{msbm10 scaled 700}

\newcommand{\ev}{{\bf e}}
\newcommand{\fv}{{\bf f}}

\newcommand{\uv}{{\bf u}}

\newcommand{\xv}{{\bf x}}
\newcommand{\yv}{{\bf y}}

\newcommand{\Dm}{{\bf D}}

\newcommand{\Id}{{\bf I}}
\newcommand{\Jm}{{\bf J}}

\newcommand{\Lm}{{\bf L}}
\newcommand{\Mm}{{\bf M}}

\newcommand{\Qm}{{\bf Q}}
\newcommand{\Rm}{{\bf R}}

\newcommand{\Um}{{\bf U}}
\newcommand{\Wm}{{\bf W}}
\newcommand{\Vm}{{\bf V}}

\newcommand{\Ec}{{\cal E}}

\newcommand{\Gc}{{\cal G}}

\newcommand{\Lc}{{\cal L}}

\newcommand{\Rc}{{\cal R}}
\newcommand{\Sc}{{\cal S}}

\newcommand{\Vc}{{\cal V}}

\newtheorem{theorem}{Theorem}
\newtheorem{definition}{Definition}
\newtheorem{prop}{Proposition}
\crefname{prop}{Proposition}{Propositions}

%% file: def_abr.tex
\newacronym{fir}{FIR}{finite-extent impulse response}
\newacronym{iir}{IIR}{infinite-extent impulse response}
\newacronym{psnr}{PSNR}{peak-signal-to-noise ratio}
\newacronym{snr}{SNR}{signal-to-noise ratio}
\newacronym{awgn}{AWGN}{additive white Gaussian noise}

\newacronym{nrmse}{NRMSE}{normalized-root-mean-square error}
\newacronym{cnn}{CNN}{convolutional neural network}
\newacronym{iid}{iid}{independent and identically distributed}
\newacronym{dft}{DFT}{discrete Fourier transform}
\newacronym{idft}{IDFT}{inverse discrete Fourier transform}
\newacronym{gft}{GFT}{graph Fourier transform}
\newacronym{igft}{IGFT}{inverse graph Fourier transform}
\newacronym{fft}{FFT}{fast Fourier transform}
\newacronym{fpga}{FPGA}{field-programmable gate array}

\newacronym{lp}{LP}{linear programming}
\newacronym{cpa}{CPA}{Chebyshev polynomial approximation}
\newacronym{ls}{LS}{least square}
\newacronym{arma}{ARMA}{autoregressive moving average}
\newacronym{wls}{WLS}{weighted least-square}
\newacronym{socp}{SOCP}{second-order cone programming}

\newacronym{sse}{SSE}{sum-of-square-error}

\newacronym{gsp}{GSP}{graph signal processing}

%% file: ICASSP_2024_main.bbl
\begin{thebibliography}{10}

\bibitem{tanaka20}
Yuichi Tanaka, Yonina~C Eldar, Antonio Ortega, and Gene Cheung,
\newblock ``Sampling signals on graphs: From theory to applications,''
\newblock {\em IEEE Signal Processing Magazine}, vol. 37, no. 6, pp. 14--30, 2020.

\bibitem{ortega22}
Antonio Ortega,
\newblock {\em Introduction to Graph Signal Processing},
\newblock Cambridge University Press, 2022.

\bibitem{anis14}
Aamir Anis, Akshay Gadde, and Antonio Ortega,
\newblock ``Towards a sampling theorem for signals on arbitrary graphs,''
\newblock in {\em 2014 IEEE International Conference on Acoustics, Speech and Signal Processing (ICASSP)}, 2014, pp. 3864--3868.

\bibitem{chen15}
Siheng Chen, Rohan Varma, Aliaksei Sandryhaila, and Jelena Kovačević,
\newblock ``Discrete signal processing on graphs: Sampling theory,''
\newblock {\em IEEE Transactions on Signal Processing}, vol. 63, no. 24, pp. 6510--6523, 2015.

\bibitem{pesenson08}
Isaac Pesenson,
\newblock ``Sampling in {Paley}-{Wiener} spaces on combinatorial graphs,''
\newblock {\em Transactions of the American Mathematical Society}, vol. 360, no. 10, pp. 5603--5627, 2008.

\bibitem{Tanaka20b_smooth}
Yuichi Tanaka and Yonina~C. Eldar,
\newblock ``Generalized sampling on graphs with subspace and smoothness priors,''
\newblock {\em IEEE Transactions on Signal Processing}, vol. 68, pp. 2272--2286, 2020.

\bibitem{Bai19}
Yuanchao Bai, Gene Cheung, Fen Wang, Xianming Liu, and Wen Gao,
\newblock ``Reconstruction-cognizant graph sampling using gershgorin disc alignment,''
\newblock in {\em 2019 IEEE International Conference on Acoustics, Speech and Signal Processing (ICASSP)}, 2019, pp. 5396--5400.

\bibitem{anis16}
A.~Anis, A.~Gadde, and A.~Ortega,
\newblock ``Efficient sampling set selection for bandlimited graph signals using graph spectral proxies,''
\newblock {\em IEEE Transactions on Signal Processing}, vol. 64, no. 14, pp. 3775--3789, July 2016.

\bibitem{Tsitsvero16}
Mikhail Tsitsvero, Sergio Barbarossa, and Paolo Di~Lorenzo,
\newblock ``Signals on graphs: Uncertainty principle and sampling,''
\newblock {\em IEEE Transactions on Signal Processing}, vol. 64, no. 18, pp. 4845--4860, 2016.

\bibitem{jayawant22}
Ajinkya Jayawant and Antonio Ortega,
\newblock ``Practical graph signal sampling with log-linear size scaling,''
\newblock {\em Signal Processing}, vol. 194, pp. 108436, 2022.

\bibitem{girault18}
Benjamin Girault, Antonio Ortega, and Shrikanth~S. Narayanan,
\newblock ``Irregularity-aware graph fourier transforms,''
\newblock {\em IEEE Transactions on Signal Processing}, vol. 66, no. 21, pp. 5746--5761, 2018.

\bibitem{pavez22}
Eduardo Pavez, Benjamin Girault, Antonio Ortega, and Philip~A. Chou,
\newblock ``Two channel filter banks on arbitrary graphs with positive semi definite variation operators,''
\newblock {\em IEEE Transactions on Signal Processing}, vol. 71, pp. 917--932, 2023.

\bibitem{jayawant23}
Ajinkya Jayawant and Antonio Ortega,
\newblock ``Towards bandwidth estimation for graph signal reconstruction,''
\newblock in {\em ICASSP 2023 - 2023 IEEE International Conference on Acoustics, Speech and Signal Processing (ICASSP)}, 2023, pp. 1--5.

\bibitem{Lee13}
Sungwon Lee and Antonio Ortega,
\newblock ``Efficient data-gathering using graph-based transform and compressed sensing for irregularly positioned sensors,''
\newblock in {\em 2013 Asia-Pacific Signal and Information Processing Association Annual Summit and Conference}, 2013, pp. 1--4.

\bibitem{perraudin2014gspbox}
Nathana{\"e}l {Perraudin}, Johan {Paratte}, David {Shuman}, Lionel {Martin}, Vassilis {Kalofolias}, Pierre {Vandergheynst}, and David~K. {Hammond},
\newblock ``{GSPBOX: A toolbox for signal processing on graphs},''
\newblock {\em ArXiv e-prints}.

\bibitem{girault17}
Benjamin Girault, Shrikanth~S. Narayanan, Antonio Ortega, Paulo Gonçalves, and Eric Fleury,
\newblock ``Grasp: A matlab toolbox for graph signal processing,''
\newblock in {\em IEEE International Conference on Acoustics, Speech and Signal Processing}, 2017, pp. 6574--6575.

\end{thebibliography}
